\documentclass[pre,onecolumn,tightenlines,12pt]{revtex4}
\usepackage{graphicx}
\usepackage{dcolumn}
\usepackage{amsmath}
\begin{document}

\title{Hyperbolic disordered ensembles of random matrices}


\author{O. Bohigas$^{1}$ and M. P. Pato$^{2}$}
 
\affiliation{
$^{1}$CNRS, Universit\'e Paris-Sud, UMR8626, \\
LPTMS, Orsay Cedex, F-91405, France \\
$^{2}$Inst\'{\i}tuto de F\'{\i}sica, Universidade de S\~{a}o Paulo\\
Caixa Postal 66318, 05314-970 S\~{a}o Paulo, S.P., Brazil}

\begin{abstract}
Using the simple procedure, recently introduced, of dividing Gaussian
matrices by a positive random variable, a family of random matrices is
generated characterized by a behavior ruled by the generalized 
hyperbolic distribution. The spectral density evolves from 
the semi-circle law to a Gaussian-like behavior while concomitantly 
the local fluctuations show a transition from the Wigner-Dyson to
the Poisson statistics. Long range statistics such as number 
variance exhibit large fluctuations typical of non-ergodic 
ensembles.

\end{abstract}

\maketitle

\section{Introduction}

In a recent paper\cite{disord}, the concept of disorder has been
introduced in random matrix theory (RMT) to denote ensembles of 
random matrices generated by superimposing an external source of 
randomness to the fluctuations of the Gaussian ensembles. 
The method generates matrices which preserve unitary invariance 
though at the price of introducing correlations among matrix elements. 
Their statistical properties are amenable to analytical derivation.
Families of ensembles recently proposed in the literature fit
into this category, for instance \cite{Bertuola,Raul} in the 
Wigner Gaussian case and, Refs. \cite{wish0,wish1,wish2} 
in the case of Wishart matrices. Disordered ensembles can be
considered as an instance of the so-called superstatistics\cite{Abul} 
aside from the fact that their density distributions are normalized differently, as
explained below.

As a consequence of having operating an extra source of
randomness, the ergodicity property of the standard ensembles, 
characterized by the equivalence between spectral averages taken 
along the spectrum and averages extracted from 
the spectra of a set of matrices is broken. This creates an ambiguity
in measuring its collective spectral fluctuations although for those which concern
individual eigenvalues such as extreme value statistics it remains well defined.
This kind of statistics has been object of many studies\cite{TW} in the last 
decades.  In this direction, it was shown how 
the behavior of the largest eigenvalue described by the Tracy-Widom 
distribution\cite{TW} is affected by disorder\cite{deform,wish1}.    

Despite the great success the RMT Gaussian ensemble had and has 
(see \cite{Mitchell} for recent reviews), it
has features of an abstract mathematical model 
such as, for example, an eigenvalue density defined in a compact support. 
Parallel to its development, there was a search for models whose 
randomness would be 
closer to physical situations. This is the case of the two-body random ensemble 
(TBRE) proposed in the context of the shell model of nuclear physics\cite{TBRE}
and of the more general $k$-body embedded Gaussian ensembles (EGE) proposed
to many-body systems\cite{EGE}.
In them, the Hamiltonian model is made randomic by taking the strength of 
the residual interaction among particles from a Gaussian distribution. 
These ensembles show as main features: correlations among matrix 
elements, Gaussian eigenvalue density and non-ergodicity\cite{Bennet}. 
Nevertheless, their spectral fluctuations are supposed to follow, under 
restricted conditions that take into account the non-ergodicity ambiguity, 
the same fluctuations pattern of the Gaussian matrices, namely the Wigner-Dyson 
statistics\cite{Meht}. Disordered ensembles may provide a simple way
to understand features of these physical motivated ensembles.

Families of disordered ensembles have, so far, been considered mostly in
association with an external randomness governed by the gamma 
distribution\cite{Bertuola,Raul}. This leads to ensembles 
in which statistics are dominated by power law behavior that implies 
strong non-ergodicity. It has also been considered the case of 
the inverse gamma distribution\cite{wish1} which contains singularities.
Our purpose here is to study an ensemble whose 
auxiliary random variable is taken from a generalized inverse 
Gaussian (GIG) which acting on the Gaussian matrices generates disordered
matrices distributed according to a generalized hyperbolic distribution (GH). 
This kind of distribution has been introduced in 1977 by 
Barndorff-Nielsen\cite{GH} and has found since then many applications  
specially in finance\cite{GHA}. 

\section{Disordered ensembles}

We start by reviewing general aspects of the formalism of the disordered ensemble 
of Ref. \cite{disord}. The idea is to construct matrices by the 
relation

\begin{equation}
H(\xi )= \frac{ H_V } {\sqrt{\xi/\bar \xi }} , \label{1}
\end{equation}
where $ H_V $ is a random matrix of dimension $N$ with joint density probability 
distribution 

\begin{equation}
P_{V} (H_V )=\frac{1}{Z_f}
\exp\left[-\frac{\beta}{2} \mbox{tr} V(H_V)\right]  \label{12}
\end{equation}
and $\xi$ is a positive random variable with a normalized 
density probability distribution $w (\xi )$ with average $ \bar{\xi}$
and variance $\sigma_{\xi}^2. $ In (\ref{12}),
$f=N+\beta N(N-1)/2$ is the number of independent matrix elements,
$\beta$ is the Dyson index $\beta=1,2,4$ and $V(x)$ is a confining potential
which makes normalization

\begin{equation}
Z_f =\int dH \exp\left[ -\frac{\beta}{2}\mbox{tr}V(H)\right]
\end{equation}
finite with respect to the measure 
$dH=\prod^{N}_{i=1}dH_{ii}\prod_{j>i}\prod^{\beta}_{k=1}\sqrt{2}dH^{k}_{ij} $.   
Rewritten as $ H_V = H (\xi) \sqrt{\xi/\bar \xi }, $
Eq. (\ref{1}) means that the joint distribution of the matrix elements of the
disordered ensemble, is
obtained from Eq. (\ref{12}) by averaging over the $\xi$ variable,
namely as

\begin{equation}
P (H  )=
\int d\xi w ( \xi)\left(\frac{\xi}{\bar{\xi}}\right)^{f/2}
\frac{1}{Z_f}\exp
\left[-\frac{\beta}{2}\mbox{tr}V\left(\sqrt{\frac{\xi}{\bar{\xi}}}H\right)  
\right] \label{9}  .
\end{equation}
Eq. (\ref{9}) translates the relation (\ref{1}) among the random 
quantities into a relation among their density distributions functions.
Despite the resemblance of Eq. (\ref{9}) with distributions
constructed in the context of the superstatistics
formalism\cite{Abul}, 
we remark that here the two distributions are integrated normalized 
while in the superstatistics case the normalization is performed 
after the integration. 

While Eq. (\ref{9}) puts in evidence the correlations among matrix
elements, the relation (\ref{1}) which is expressed 
in terms of the random quantities themselves,  
gives a clearer picture of
the model and, at the same time, provides an efficient algorithm to
generate the disordered matrices. We also remark that in \cite{disord}, 
an alternative procedure has been described in which matrices of the 
ensemble are generated taking into account correlations 
among their elements (see next section).

Turning now to eigenvalues and eigenvectors, we observe that by
diagonalizing the matrices in  (\ref{1}), the
relation 

\begin{equation}
D(\xi )= \frac{ D_V } {\sqrt{\xi/\bar \xi }}  \label{1b}
\end{equation}
is obtained, where $D(\xi)$ and $D_V$ are diagonal matrices whose 
elements are the respective eigenvalues of $H(\xi)$ and $H_V$. 
Rewriting, as done with  Eq. (\ref{1}), Eq. (\ref{1b}) as  
$ D_V = D (\xi) \sqrt{\xi/\bar \xi }, $ the relation 

\begin{equation}
P\left( E_{1},...E_{N}\right ) =
\int d\xi w \left( \xi \right) \left(\frac{\xi}{\bar{\xi}}\right)
^{\frac{N}{2}} P_V \left( x_{1},x_2,...x_{N}\right)  
\label{15}
\end{equation}
is derived, where $x_i = \sqrt{\xi/\bar{\xi}}E_i $ and

\begin{equation}
P_V \left( x_{1},...x_{N}\right ) =
K_{N}^{-1}\exp\left[-\frac{\beta}{2}\sum_{k=1}^{N} V(x_{k})\right]
J\left( x_{1},x_2,...x_{N}\right) .  \label{15c}
\end{equation}
In  (\ref{15c}), $K_{N}$ is a normalization constant
and $J\left( x_{1},x_2,...x_{N}\right)$ is the eigenvalue part of 
the Jacobian of the transformation from matrix elements to eigenvalues 
and eigenvectors. From (\ref{15}), 
statistical measures of the generalized family can be calculated by 
weighting, with the $w(\xi)$ distribution, the corresponding measures 
of the original ensemble with the eigenvalues multiplied
by the factor $\xi/\bar{\xi}$. 

Considering particular choices of the distribution
$w(\xi)$, the most used so far both, in the Gaussian and 
the Wishart cases, was the gamma distribution  

\begin{equation}
w(\xi)=
\exp(-\xi) \xi^{\bar{\xi} -1} /\Gamma(\bar{\xi})  \label{18}
\end{equation} 
with variance $\sigma_{\xi}=\sqrt{\bar{\xi}}$ .  Considering the 
case of the Wigner ensemble in which $V(H)=H^2,$ when (\ref{18}) 
is substituted into  (\ref{9}), the integrals are readily performed 
to give the joint distribution

\begin{equation}
P(H;\bar{\xi} )=\frac{1}{Z_f}\frac
{\Gamma \left( \bar{\xi}+f/2\right)}{\Gamma \left( \bar{\xi} \right) }
\left(1+\frac{\beta}{2\bar{\xi}}
\mbox{tr} V(H) \right) ^{-\bar{\xi}-f/2}  \label{22} .
\end{equation}
To illustrate the power-law behavior induced by the above gamma
distribution, take out of the $f$ independent elements, the density 
distribution of a given one
denoted by $h$ (see next section for the definition of the $h$
variable), after integrating over the $f-1$ others,
we obtain

\begin{equation}
p(h;\bar{\xi})=\left(\frac{\beta}{2\pi\bar{\xi}}
\right)^{\frac{1}{2}}\frac{\Gamma \left(\bar{\xi} +1/2 \right) }
{\Gamma \left( \bar{\xi}\right)}
\left(1+\frac{\beta}{2\bar{\xi}}
h^{2}\right) ^{-\bar{\xi}-1/2}  \label{280} .
\end{equation}
Since for large $\left|h\right|,$ 
$p_{\beta}(h;\bar{\xi}) \sim 1/\left|h\right|^{2\bar{\xi}+1},$ 
the distribution (\ref{280}) does not have moments of order superior to
$2\bar{\xi}.$ This fact makes the value $\bar{\xi} =1$ critical since
below it (\ref{280}) does not have a second moment\cite{Bertuola}. 
We remark that, apart from the lack of independence, 
the marginal distribution of the matrix 
elements has the same kind of distribution, namely one with an 
asymptotic power-law behavior, as the ones of the ensemble 
of L\'{e}vy matrices\cite{Cizeau}.

Another distribution considered in \cite{wish1} was the 
inverse distribution obtained by making $\xi=1/\nu$ which leads to 

\begin{equation}
w(\nu)=\frac{\bar{\nu}}{\Gamma(1/\bar{\nu})}
\exp\left(-\frac{1}{\nu}\right) \nu^{-\frac{1}{\nu} -2} .  \label{18b}
\end{equation} 
Again, the integral in (\ref{9}) can be performed to give

\begin{equation}
P(H;\bar{\nu} )=\frac{2}{Z_f \Gamma(1/\bar{\nu})}
\left(\sqrt{\frac{\bar{\nu}}{2}\beta\mbox{tr} V(H)}\right)^{1+\frac{1}{\bar{\nu}}-f/2}
K_{f/2-1/\bar{\nu}-1}\left(\sqrt{\frac{\bar{\nu}}{2}\beta\mbox{tr} V(H)}
\right)   \label{22b} 
\end{equation}
which shows, as a drawback, the presence of a singularity at the origin.

Our present purpose is to study the model generated by taking the disorder  
variable out of the generalized inverse Gaussian   

\begin{equation}
w(a,b,\lambda;\xi)= \frac{(b/a)^{\lambda/2}}
{2K_{\lambda}(\sqrt{ab})}\xi^{\lambda-1}\exp
\left[-\frac{1}{2}(\frac{a}{\xi}+b\xi)\right], \label{28}
\end{equation}
where $K_{\lambda}(x)$ is the modified Bessel function of the third
kind. This is a three parameter probability distribution where $a$ and
$b$ are positive and $\lambda$ is real. It contains the above distributions 
(\ref{18}) and (\ref{18b}) as special cases. In fact, in the limit 
$a\rightarrow 0$, apart from scaling, the gamma 
distribution, Eq. (\ref{18}), is recovered while in the limit  $b\rightarrow 0$
the inverse distribution is obtained. The GIG first moment is

\begin{equation}
\bar{\xi}=\sqrt{\frac{a}{b}}\frac{K_{\lambda+1}(\sqrt{ab})}
{K_{\lambda}(\sqrt{ab})}
\end{equation}
and its variance is

\begin{equation}
\sigma_{\xi}^2=\left(\frac{b}{a}\right)\left[\frac{K_{\lambda+2}(\sqrt{ab})}
{K_{\lambda}(\sqrt{ab})}-\frac{K^2_{\lambda+1}(\sqrt{ab})}
{K^2_{\lambda}(\sqrt{ab})}\right]  .
\end{equation}
By taking the ratio

\begin{equation}
\frac{\bar{\xi}}{\sigma_{\xi}}=\frac{b}{a}\sqrt{\frac{K_{\lambda+2}(\sqrt{ab})
K_{\lambda}(\sqrt{ab})}
{K_{\lambda+1}^2(\sqrt{ab})}-1} ,  \label{28d}
\end{equation}
we have a better parameter to understand the effect of the disorder
generated by the GIG
distribution.  First we remark that, by changing 
in (\ref{28d})  the sign of $\lambda, $ the symmetry property 
$K_{-\nu}(x)=K_{\nu}(x)$ of the Bessel functions, makes the above ratio 
symmetric with respect to the value $\lambda =-1.$  Secondly,
for large orders,  the Bessel functions can be
approximated by the dominant term
of their uniform approximation\cite{Abram}

\begin{equation}
K_{\nu}(\nu z)\sim \sqrt{\frac{\pi}{2\nu}}
\frac{\exp(-\mu\nu)}{(1+z^2)^{1/4}}  ,
\end{equation}
where

\begin{equation}
\mu=\sqrt{1+z^2} + \ln\frac{z}{1+\sqrt{1+z^2}} .
\end{equation}
Using this,  the Bessel functions
in (\ref{28d}) can be approximated for large $\lambda$ as

\begin{equation}
K_{\lambda}(\sqrt{ab})\sim
\sqrt{\frac{\pi}{2\lambda}}\left(\frac{\lambda}{\sqrt{ab}}\right)^{\lambda}
\exp(-\lambda),
\end{equation}
where, in presence of
one, the square of the term $\sqrt{ab}/\lambda$ has been neglected. 
As a consequence, with $a=b,$  the ratio (\ref{28d}) becomes
practically constant when $\lambda$ increases as indeed can be
seen in Fig. 1 . 
On the other hand, by fixing $\lambda$ and the product $ab$, 
the GIG also becomes more localized as the ratio $b/a$ decreases.
as illustrated in the same Fig. 1. From
this analysis, we expect larger disorder for small values of the 
parameter $\lambda$ or large values of the ratio $b/a,$ with the 
value  $\lambda =-1$ playing a special role in this 
generalized family.  

\section{Hyperbolic disordered Gaussian ensemble}

The Wigner ensemble is obtained by making $V(H)=H^2$ with $H$ a Hermitian matrix.
The number of independent
matrix elements is of course $f=N+\beta N(N-1)/2$ and it is convenient to consider
the set of $f$ independent elements denoted by $h_n$ with
$n=1,2,3,...,f.$ They are defined such that first $N$ ones are equal
to the diagonal elements, that is, $h_i =H_{ii}.$ The remain ones,
the off-diagonal elements, are scaled as $h_n =\sqrt{2}H_{ij}^{k}$
with $n=N+1,N+2,...,f.$ 
In terms of these $f$ independent variables, the joint distribution,
Eq. (\ref{9}), together with Eq. (\ref{28}) becomes

\begin{equation}
P_f (h_1,...,h_f)= \frac{(b/a)^{\lambda/2}}
{2K_{\lambda}(\sqrt{ab})}\int^{\infty}_{0}d\xi\xi^{\lambda-1}
\left(\frac{\beta\xi}{2\pi\bar{\xi}} \right)^{f/2}\exp
\left[-\frac{1}{2}(\frac{a}{\xi}+b\xi)-\sum_{n=1}^{f}
\frac{\beta\xi h^2_n}{2\bar{\xi}}\right]  .   \label{9g}
\end{equation}
Each $h_n$ variable is treated equally in (\ref{9g}) and that is the
reason to introduce this set of variables in which the inconvenient
factor of $\sqrt{2}$ multiplying the off-diagonal ones has been absorbed in
the new variable. The integration over the extra variable $\xi$ makes
the distribution in (\ref{9g}) a correlated distribution of the set of
variables, but the above distribution has the
important property that removing by integration any one variable, the
distribution of the remnant others has the same form, namely,we have 

\begin{equation} 
P_{f-1} (h_1,...,h_{f-1}) =\int dh_f P_f (h_1,...,h_f) .
\end{equation}
This property is called scale-invariant occupancy of phase
space\cite{Sato}. Of course, it is a manifestation of the
uncorrelation among the matrix elements of the original ensemble. 
In this way, one can expect this property to hold in general for 
the disordered ensembles. On the other side, it is not expected to
hold for generalizations of distributions
constructed using the strict superstatistics recipe. 
    
As it occurred with the gamma distribution and its inverse, after 
substituting (\ref{28}) in 
(\ref{9}), the integrals are performed and we find  the joint 
distribution of  matrix elements and the distribution of a generic
matrix element $h$ to be given, respectively, by the generalized 
hyperbolic distributions

\begin{equation}
P(a,b,\lambda;H)=\left(\frac{\sqrt{a/b}}{2\pi\bar{\xi}}\right)^{f/2}
\frac{1}{\left(1+\frac{1}{b\bar{\xi}}\sum_{i=1}^{f} h_i^2\right)^{\lambda_f/2}}
\frac{K_{\lambda_f}\left[\sqrt{ab\left(1+\frac{1}{b\bar{\xi}}\sum_{i=1}^{f} h_i^2
\right)}\right]}{K_{\lambda}(\sqrt{ab})} , \label{28a}
\end{equation}
and

\begin{equation}                               
p(a,b,\lambda;h)=\frac{(a/b)^{1/4}}{\sqrt{2\pi\bar{\xi}}}                                 
\frac{1}{\left(1+\frac{h^2}{b\bar{\xi}}\right)^{\lambda_1/2}}                          
\frac{K_{\lambda_1}\left[\sqrt{ab\left(1+\frac{h^2}{b\bar{\xi}}                             
\right)}\right]}{K_{\lambda}(\sqrt{ab})} ,  \label{28b}                                     
\end{equation}
where with $1\le k \le f,$ $\lambda_k =\lambda+k/2.$ The
scale-invariant property has been used to write Eq. (\ref{28b}). 
Comparing these GH equations with the correspondent equations 
for the gamma disorder, we see that the GH ones contain
in themselves the gamma ones but multiplied by the Bessel function
that decays exponentially for large values of $|h|.$ This leads to 
a more regular behavior as becomes clear from the expression 

\begin{equation}
\overline{h^n}=\frac{\left(2\bar{\xi}b/a\right)^{n/2}}{K_{\lambda}}
\frac{\Gamma[(n+1)/2]}{\sqrt{\pi}}K_{\lambda-n/2}\left(\sqrt{ab}\right).
\end{equation}
for the  moment of arbitrary $n$th order of (\ref{28b}).
From the symmetry property of the Bessel function, i.e. 
$K_{-\nu}(x)=K_{\nu}(x)$ we conclude that (\ref{28b}) has all moments,
resulting in milder fluctuations.

Consider the identity

\begin{equation}
P_f (h_1,...,h_f)=\frac{P_f (h_1,...,h_f)}{P_f (h_1,...,h_{f-1})}...
\frac{P_f (h_1,...,h_k)}{P_f (h_1,...,h_{k-1})}...
\frac{P_2 (h_1,h_2)}{P_1(h_1)}P_1(h_1)
\end{equation}
in which by definition each ratio gives the 
conditional probability $p(h_k|h_1,...,h_{k-1})$ of the last element 
$h_k$ in the numerator argument 
once the preceding $k-1$ ones have been determined. Using the property of 
scale-invariance, this probability can be written in terms of the Bessel 
functions as 
  
\begin{equation}                                                                            
p(h_k|h_1,h_2,...,h_{k-1})=\frac{(a/b)^{1/4}}{\sqrt{2\pi\bar{\xi}}}
\frac{\left(1+\frac{1}{b\bar{\xi}}\sum_{1}^{k-1}h_{i}^2 \right)^{\frac{\lambda_{k-1}}{2}}}
{\left(1+\frac{1}{b\bar{\xi}}\sum_{1}^{k}h_{i}^2 \right)^{\frac{\lambda_{k}}{2}}}
\frac{K_{\lambda_k}\left[\sqrt{ab\left(1+\frac{1}{b\bar{\xi}}\sum_{1}^{k}h_{i}^2
\right)}\right]}
{K_{\lambda_k}\left[\sqrt{ab\left(1+\frac{1}{b\bar{\xi}}\sum_{1}^{k-1}h_{i}^2
\right)}\right]}   .               
\end{equation}
A more compact expression is obtained doing the following

\begin{equation}
1+\frac{1}{b\bar{\xi}}\sum_{i=1}^{k-1}h_{i}^2 +\frac{h^2_k}{b\bar{\xi}}=
\left(1+\frac{1}{b\bar{\xi}}\sum_{i=1}^{k-1}h_{i}^2 \right)\left[1+\frac{h_{k}^{2}}
{b\bar{\xi}\left(1+\frac{1}{b\bar{\xi}}\sum_{i=1}^{k-1}h_{i}^2 \right)}\right]=
\frac{b_{k-1}}{b}\left(1+\frac{h_{k}^{2}}{b_{k-1}\bar{\xi}}\right) ,
\end{equation}
where

\begin{equation} 
b_k =b\left(1+\frac{1}{b\bar{\xi}}\sum_{i=1}^{k}h^2_i\right) ,
\end{equation}
with $b_0=b.$ With these definitions the probability distribution of
one element $h_k$ once $k-1$ other
ones have already been sorted is  

\begin{equation}                                                                            
p(h_k|h_1,h_2,...,h_{k-1})=\frac{(a/b_{k-1})^{1/4}}{\sqrt{2\pi\bar{\xi}}}
\frac{1}{\left(1+\frac{h_k^2}{b_{k-1}\bar{\xi}}\right)^{\lambda_k/2}}
\frac{K_{\lambda_k}\left[\sqrt{ab\left(1+\frac{h_k^2}{b_{k-1}\bar{\xi}}
\right)}\right]}{K_{\lambda_{k-1}}(\sqrt{ab_{k-1}})}   ,               
\end{equation}
With $k$ running from $2$ to $f,$ these equations form a set of univariate
distributions which can sequentially be used to generate a matrix of the ensemble
in which correlations among elements are taken into account.

Integrating (\ref{15}) over all eigenvalues but one and multiplying 
by $N,$ the eigenvalue density is expressed in terms of 
the Wigner's semi-circle law density $\rho_G(E)=\sqrt{2N-E^2}/\pi$ 
of the Gaussian ensemble\cite{Meht}  as

\begin{equation}
\rho \left( E\right) =\frac{1} {\pi}\int_{0}^{\xi_{max}} 
d\xi w( \xi) \left(\frac{\xi}{\bar{\xi}}\right)^{1/2}\sqrt{2N- 
\frac{\xi}{\bar{\xi}}E^2 }. \label{126}
\end{equation}
where $\xi_{max}=2N\bar{\xi}/E^2$. From (\ref{126}), 
we find that at the origin

\begin{equation}
\rho (0)=\frac{\sqrt{2N}}{\pi}\frac{\bar{\sqrt{\xi}}}{\sqrt{\bar{\xi}}}
=\rho_G (0)\frac{\bar{\sqrt{\xi}}}{\sqrt{\bar{\xi}}} \label{31q}
\end{equation}
and, since the ratio between the two averages is less than one, the disordered 
density is, at the origin, smaller than the semi-circle value. 
When $|E|\rightarrow\infty,$ $\xi_{max}\rightarrow 0$ and therefore the interval 
of integration in (\ref{126}) collapses and a crude approximation
to the integral is

\begin{equation}
\rho(E)\sim \exp\left(-\frac{aE^2}{2N\bar{\xi}}\right) E^{-2\lambda-3} 
\end{equation}
which shows that the density has Gaussian tails. In Fig.2, it is shown
that with the parameter $\lambda$ fixed at the value $-1,$ 
the density undergoes a transition from the semi-circle to a
Gaussian-like shape which becomes more and more deformed as $a=b$
decreases.

By integrating the density, Eq. (\ref{126}), from the origin to a value $E$, 
we obtain the function

\begin{equation}
N (E)=\frac{N}{2}\left(1-\int_0^{\xi_{max}}d\xi w(\xi)\left[1-\frac{2}{N}
N_G\left(\sqrt{\frac{\xi}{\bar{\xi}}}E \right)\right]\right) ,
\end{equation}
where

\begin{equation}
N_G(E)=\frac{N}{2}\left(\arcsin\frac{E}{\sqrt{2N}}+\frac{E}{\sqrt{2N}}\sqrt 
{1-\frac{E^2}{2N}}\right) .  
\end{equation}
The functions $N(E)$ and $N_G (E)$ count the average number of eigenvalues 
in the interval $(0,E)$ for the the disordered and the 
Gaussian ensembles, respectively.

To measure spectral fluctuations, the dependence on the density is 
removed by the transformation $x=N(E)$ which  maps the actual spectrum 
into a new one with unit density. Starting with short range correlations, 
we define the spacing function that gives 
the probability $E(0,s)$ that the interval $(-\frac{s}{2},\frac{s}{2})$ 
is empty. To calculate this function we integrate Eq. (\ref{15}) 
over all eigenvalues outside the interval the 
$(-\frac{\theta}{2},\frac{\theta}{2})$ to obtain 
with $s=2N(\frac{\theta}{2})$ 

\begin{equation}
E(0,s)=\int_0^{\infty}d\xi w(\xi) E_G\left[0,2N_G\left(
\sqrt{\frac{\xi}{\bar{\xi}}}\frac{\theta}{2}\right)\right], \label{108s}
\end{equation} 
where for the Gaussian spacing we use the Wigner surmise for the GOE
case ($\beta=1$)

\begin{equation}
E_G (0,s)=\mbox{erfc} (\sqrt{\frac{\pi }{4}}s )  .
\end{equation}

From this function the nearest neighbor distribution (NND) is derived as
$p(s)=\frac{d^2 E(0,s)}{ds^2}$ which gives 

\begin{equation}
p(s)=\left\{ 
\begin{array}{rl}
-\frac{\rho^{\prime}(\theta/2)E^{\prime}(0,s)}{2\rho^3 (\theta/2)}
+\frac{1}{2\rho^2 (\theta/2)}\int_0^{\xi_m}d\xi w(\xi)(\frac{\xi}{\bar{\xi}})
\rho_G^{\prime}\left(\sqrt{\frac{\xi}{\bar{\xi}}}\frac{\theta}{2}\right)
E^{\prime}_G\left[0,2N_G(\sqrt{\frac{\xi}{\bar{\xi}}}\frac{\theta}{2})\right]\\
+\frac{1}{\rho^2 (\theta/2)}\int_0^{\xi_m}d\xi w(\xi)(\frac{\xi}{\bar{\xi}})
\rho_G^{2}\left(\sqrt{\frac{\xi}{\bar{\xi}}}\frac{\theta}{2}\right)
p_G\left[0,2N_G(\sqrt{\frac{\xi}{\bar{\xi}}}\frac{\theta}{2})\right]
\end{array}
\right.
\end{equation}

The above equations are exact and the spacings  presented below
in Fig. 3 were calculated with them. To understand these results 
we derive approximated spacings considering that for large $N,$ 
the two counting functions, $N(E)$ and $N_G (E)$ can be replaced, 
at the center of the spectrum, by $N(E)=\rho(0)E$ and 
$N_G (E)=\rho_G (0)E.$ With these approximations, Eq. (\ref{108s}) 
becomes

\begin{equation}
E(0,s)=\frac{(b/a)^{\frac{\lambda}{2}}}{2K_{\lambda}(\sqrt{ab})}
\int_0^{\infty}d\xi \exp\left[-\frac{1}{2}(\frac{a}{\xi}+b\xi)\right] 
\xi^{\lambda-1}\mbox{erfc}\left[\frac{\sqrt{\xi\pi}\rho_{W}(0)}
{2\sqrt{\bar{\xi}}\rho(0)}s\right]
\end{equation}
and its derivative $F(0,s)=-dE(0,s)/ds$

\begin{equation}
F(0,s)=\frac{(b/a)^{\frac{\lambda}{2}}}{2K_{\lambda}(\sqrt{ab})}
\frac{\rho_{W}(0)}{\rho(0)\sqrt{\bar{\xi}}}
\int_0^{\infty}d\xi e^{-\frac{1}{2}\left[\frac{a}{\xi}+
\left(b+\frac{\pi\rho_{W}^{2}(0)s^2}{2\bar{\xi}\rho^{2}(0)}\right)\xi \right]}
\xi^{\lambda-\frac{1}{2}} . \label{31h}
\end{equation}
We remark that using Eq. (\ref{31q}) it can be verified that the condition 
$F(0,0)=1$ is satisfied.
We can go further by expressing Eq. (\ref{31h}) in terms of Bessel functions as

\begin{equation}
F(0,s)=\frac{(b/a)^{\frac{\lambda}{2}}}{K_{\lambda}(\sqrt{ab})}
\frac{\rho_{W}(0)}{\rho(0)\sqrt{\bar{\xi}}}
\frac{K_{\lambda_1}\left[\sqrt{ab\alpha(s)}\right]}
{\left[\alpha(s)\right]^{\lambda_{1/2}}} ,
\end{equation}
where 

\begin{equation}
\alpha(s)=1+\frac{\pi\rho^2(0)}{2b\rho^2_G(0)\bar{\xi}}s^2  .
\end{equation}
Finally, in the same approximation the NND has the
expression

\begin{equation}
p(s)=\frac{K^2_{\lambda}(\sqrt{ab})}{K^3_{\lambda_1}(\sqrt{ab})}
\frac{K_{\lambda_3}\left[\sqrt{ab\alpha(s)}\right]}
{\left[\alpha(s)\right]^{\lambda_3/2}}\frac{\pi}{2}s . \label{35}
\end{equation}
Replacing now the Bessel function by its asymptotic behavior, assuming that 
as a function of $s,$ its argument is large,
we find the NND decays as 
\begin{equation}
p(s) \sim \exp\left(-\sqrt{ab\left[1+\frac{\pi\rho^2(0)}
{2b\rho^2_G(0)\bar{\xi}}s^2\right] }\right) .   \label{35k}
\end{equation}
This decaying can present two limiting situations, first, if $a$ and $b$ increases,
the second term inside the square root becomes smaller than one, in such a
way that  a Gaussian decay is obtained by expanding the square root. In the 
second situation, $b$ decreases and makes the second term much greater than one
such that it becomes the dominant term leading to an exponential decay.   

Therefore, as a function of its parameters, this model constitutes a family which
locally describes intermediate cases between the Wigner-Dyson and the 
Poisson statistics. This is illustrated in Fig. 3 where 
the cumulative NND, $F(s)=1-F(0,s),$ is plotted by the indicated values of 
the parameters. We see that as the density goes from the semi-circle 
to the Gaussian-like shape, concomitantly the spacing moves from the Wigner 
surmise to the Poisson distribution $1-\exp(-s).$ Further, at an intermediate 
point of the transition, which coincides with the density approaching a Gaussian 
shape, the cumulative NND approaches the so-called semi-Poisson distribution 
given by $ 1-(1+2s)\exp(-2s)$\cite{Bog}.

A statistics that measures spectral long range correlations 
and, also estimates non-ergodicity\cite{disord,Pandey}, is the variance 
$ \Sigma^2 (L) $ of the number of eigenvalues in the interval 
$[ -\theta/2,  \theta/2]$ with $L=2N(\theta/2)$.  
In \cite{disord} it is shown that this variance is given by 
 
\begin{equation}
\Sigma^2 \left( L \right) =\int d\xi w
\left(\xi \right) \left[
\Sigma^2_G \left(2N_G(\sqrt{\frac{\xi}{\bar{\xi}}}\frac{\theta}{2})\right) - 
2N_G\left(\sqrt{\frac{\xi}{\bar{\xi}}}\frac{\theta}{2} \right) + 4N_G ^2 
\left(\sqrt{\frac{\xi}{\bar{\xi}}}\frac{\theta}{2} \right)  \right]+ 
L - L^2 .\label{37}
\end{equation}
where $ \Sigma^2_G  $ is Gaussian number variance. In the limit
of large spectra, we again use the linear approximations of
the two counting functions to obtain

\begin{equation}
\Sigma^2 \left( L \right) =
=\Sigma^2_G \left[\frac{\rho_G(0)}{\rho(0)}L\right] 
+ \left[\frac{\rho_G(0)}{\rho(0)}-1 \right] L^2 ,\label{38}
\end{equation}
where using the fact that $\Sigma^2_G (x)$ is a logarithmic smooth 
function of its argument, the integral was asymptotically performed.
Eq. (\ref{38}) shows that the effect of disorder in the number variance
statistics is twofold: (i) it enhances it by rescaling the argument of
the Gaussian expression and (ii) it introduces a superpoissonian quadratic 
term that affects large intervals. These effects are illustrated in
Fig. 4, in which Eq. (\ref{38}) is calculated with parameters 
values which give a density and a spacing distribution closest to the 
Wigner-Dyson's ones in Fig.2 and 3.

To study the behavior of the largest eigenvalues in the limit 
of large matrix size $N,$ one introduces the scaled variable

\begin{equation}
s(E)=\sqrt{2}N^{2/3}\left[\frac{E}{\sqrt{2N}}-1\right].
\end{equation}
in terms of which 
the probability $E_{G,\beta}\left(0,s\right)$ 
that the infinite interval $(s,\infty)$ is empty is known
for the three symmetry classes\cite{TW}. For the unitary
class, $\beta=2$

\begin{equation}
E_{G,2}(s) =\exp\left[-\int_{s}^{\infty}
(x-s)q^{2}(x)dx\right]
\label{148}
\end{equation}
where $q(s)$ satisfies the  Painlev\'{e} II equation

\begin{equation}
q^{\prime\prime} = s q +2q^3
\label{149}
\end{equation}
with boundary condition 

\begin{equation}
q(s)\sim \mbox{Ai}(s) \mbox{  when  } s 
\rightarrow \infty, \label{418}
\end{equation} 
where $ \mbox{Ai}(s)$ is the Airy function.
For the orthogonal ($\beta=1$) and symplectic ($\beta=4$) classes
the probabilities are given, respectively,  by

\begin{equation}
\left[E_{G,1}(s)\right]^2 =E_{2} (0,s)\exp\left[\mu (s)\right]
\end{equation}
and

\begin{equation}
\left[E_{G,4}(s)\right]^2 =E_{2} (0,s)
\cosh^{2} \frac{\mu (s)}{2}
\end{equation}
where 

\begin{equation}
\mu (s) =\int_{s}^{\infty} q(x)dx. \label{444}
\end{equation}

The above equations give a complete description of the fluctuations 
of the eigenvalues at the edge of the spectra of the Gaussian
ensembles. When perturbed, it has been shown\cite{deform} 
that for the three symmetry classes the above probabilities 
modify to

\begin{equation}
E_{\beta}\left(\lambda_{max} < t\right)=
\int d\xi w (\xi )
E_{G,\beta}\left[S(\xi,t)\right] 
\label{146}
\end{equation}
with the argument of $S(\xi,t)$ obtained by plugging in 
the above $\xi$-variance namely

\begin{equation}
S(\xi,t)=\sqrt{2}N^{2/3}\left[\frac{t}{\sqrt{2N}}\sqrt{\frac{\xi} 
{\bar{\xi}}}-1\right].
\label{147}
\end{equation}

Equations (\ref{146}) and (\ref{147})  
give a complete analytical description of the
behavior of the largest eigenvalue once the function $w(\xi)$ is
chosen. In \cite{deform},  asymptotic results, when the limit 
$N\rightarrow \infty$ is taken, have been derived without specifying  
$w(\xi).$ To do this, it was assumed that the localization 
of $w(\xi),$ given by the ratio $\sigma_w /\bar{\xi}$ was 
dependent on the matrix size $N.$
Considering the distribution $w(\xi)$ independent of $N$ 
keeping the ratio $ \frac{t}{\sqrt{2N}}  $ fixed
when the matrix size $N$ increases, for the three invariant ensembles, 
the function $E_{G,\beta}$ becomes a step function 
centered at $\xi=2N\bar{\xi}/t^2.$ Therefore, in this regime, 
the probability distribution for the largest eigenvalue converges to

\begin{equation}
E_{\beta}\left(\lambda_{max} < t\right)=\int_{2N\bar{\xi}/t^2}^{\infty}d\xi 
w\left(\xi \right)  
\label{164}
\end{equation}
with density

\begin{equation}
\frac{dE_{\beta}\left( t\right)}{dt} =
\frac{4N \bar{\xi}w(2N\bar{\xi}/t^2)}
{ t^{3}} . \label{605}
\end{equation}
that shows a GIG distribution for the square of the largest eigenvalue. 

\section{Hyperbolic disordered Wishart matrices}

Taking now $V= X^{\dagger}X $ with $X$ being a rectangular matrix $X$ 
of size ($M$x$N$), the Wishart
ensemble is defined by the joint density distribution

\begin{equation}
P_{W} (X )=\left(\frac{\beta}{2\pi}\right)^{f/2}
\exp\left(-\frac{\beta}{2}\mbox{tr}(X^{\dagger}X) \right) \label{101},  
\end{equation}
where $f=\beta MN.$ From (\ref{101}), the elements of $X$ are Gaussian 
distributed. The eigenvalues of the random matrix  
$ V=X^{\dagger}X$ have joint density distribution

\begin{equation}
P_W (x_1,x_2,...,x_N)=K_N \exp(-\frac{\beta}{2}\sum_{k=1}^N x_k)
\prod_{i=1}^{N} x_i^{\frac{\beta}{2}(1+M-N)-1}
\prod_{j>i}\mid x_j-x_i\mid^\beta .
\end{equation}
This ensemble was introduced by the statistician J. 
Wishart\cite{Wishart} and plays an important r\^{o}le 
in statistical analysis\cite{Wilks} where $V= X^{\dagger}X $ is 
a covariant matrix. More recently, it appeared
associated in the chiral random ensemble\cite{Verba} when the square of its
eigenvalue is taken.

It can be shown that, for large matrices, the density of 
these eigenvalues approaches the Marchenko-Pastur density\cite{Pastur}

\begin{equation}
\rho_{MP} (x)=\frac{1}{2\pi x}
\sqrt{(x_{+}-x)(x-x_{-})}, 
\end{equation}
in which, with $c=\sqrt{\frac{M}{N}}$, $ x_{\pm}=N(c\pm 1)^2.$
The counting function obtained integrating this density is 

\begin{equation}
N_W (x) =\frac{1}{4\pi}\left[
\begin{array}{rl}
-4\sqrt{x_{-}x_{+}}\arctan \sqrt{\frac{x_{+}(x-x_{-})}{x_{-}(x_{+}-x)}}    
+(x_{+}+x_{-})\arccos\left(\frac{x_{+}+x_{-}-2x}
{x_{+}-x_{-}}\right)                            \\
+ 2\sqrt{(x_{+}-x)(x-x_{-})}      
\end{array}  
\right.                         .
\end{equation}

As in the Wigner case, disorder is introduced in the ensemble by defining 
new matrices through the relation

\begin{equation}
X(\xi )= \frac{ X_W } {\sqrt{\xi/\bar \xi }} , \label{102}
\end{equation}
which replaced in (\ref{101}) leads to an ensemble with joint density
distribution of matrix elements

\begin{equation}
P (X ) =
\int d\xi w ( \xi)
\left(\frac{\beta\xi}{2\pi\bar{\xi}}\right)^{f/2}
\exp\left(-\frac{\beta\xi}{\bar{2\xi}}\mbox{tr}(X^{\dagger}X) \right). \label{104}
\end{equation}
After substituting (\ref{28}) in (\ref{104}), as it occurred in the Wigner case,
integrals can be performed and we get 

\begin{equation}
P(a,b,\lambda;H)=\left(\frac{\sqrt{a/b}}{2\pi\bar{\xi}}\right)^{f/2}
\frac{1}{\left(1+\frac{\mbox {tr}(X^{\dagger}X) }{b\bar{\xi}}\right)^{\lambda_f/2}}
\frac{K_{\lambda_f}\left[\sqrt{ab\left(1+\frac{\mbox {tr}(X^{\dagger}X)}{b\bar{\xi}}
\right)}\right]}{K_{\lambda}(\sqrt{ab})} .
\end{equation}

The joint distribution of eigenvalues is given by Eq. (\ref{15}) with $P_V$ 
replaced by $P_W$. 
Eigenvalue measures are therefore obtained by averaging those of the
Wishart matrices and, for instance, the eigenvalue density is given by  

\begin{equation}
\rho ( x ) =\frac{1} {2\pi x}\int_{\xi_{-}}^{\xi_{+}} 
d\xi w( \xi) \left(\frac{\xi}{\bar{\xi}}\right)^{1/2}
\sqrt{ \left(x_{+}-x\sqrt{\frac{\xi}{\bar{\xi}}}\right)
\left(x\sqrt{\frac{\xi}{\bar{\xi}}}-x_{-}\right) }, \label{106}
\end{equation}
where

\begin{equation}
\xi_{\pm} =\bar{\xi} \left( \frac{  x_{\pm}}{x}\right)^2 .
\end{equation}
Of course, the above density extends beyond the Marchenko-Pastur
limits $x_\pm$ as can be seen in Fig. 5. By taking the limit 
$x\rightarrow\infty,$ the integration interval collapses and a 
crude approximation to the integral gives the exponential decay

\begin{equation} 
\rho(x)\sim \exp \left[ -\frac {a x^2}{2\bar \xi(x_{+}^2 +x_{-}^2)}\right]/x.  
\end{equation}
For the counting function we find after integrating the density

\begin{equation}
N (x)=\int_{\xi_{-}}^{\xi_{+}} d\xi w(\xi)
N_W \left(\sqrt{\frac{\xi}{\bar{\xi}}}x \right) 
+N\left[1-\int_{0}^{\xi_{+}}d\xi w(\xi)\right] \label{106b} .
\end{equation}

The Wishart ensemble belongs to a class of random ensembles whose spectral
fluctuations are derived from a general formalism based on orthogonal 
polynomials, the Laguerre ones, and share the same universal statistiscal 
properties\cite{Damgaard}. This universality in the case of the Wigner-Dyson 
statistics is manifested after mapping the eigenvalues into variables with density one.
We consider two spacing functions which give the probability of an 
interval to be empty: one at the bulk and the other at the inferior extreme
of the spectrum.

At the bulk, we take the interval $(\bar{x}-\theta/2,\bar{x}+\theta/2)$ where
$\bar{x}=M$ is the average position  of the Marchenko-Pastur density. 
The spacing or gap function after introducing disorder is related to
the unperturbed one by  

\begin{equation}
E(0,s)=\int_0^{\infty}d\xi w(\xi) \left(
E_G\left[0,N_W\left(\sqrt{\frac{\xi}{\bar{\xi}}}(M+\frac{\theta}{2})\right)
\right]-
E_G\left[0,N_W\left(\sqrt{\frac{\xi}{\bar{\xi}}}(M-\frac{\theta}{2})\right)
\right] \right), \label{108q}
\end{equation}    
where the spacing $s$ is calculated with Eq. (\ref{106b}) as 

\begin{equation}
s= N\left(x=M+\frac{\theta}{2}\right) -  N\left(x=M-\frac{\theta}{2}\right).
\end{equation}
Comparing (\ref{108q}) and (\ref{108s}), we conclude that these
spacings have the same behavior.

To study the behavior of the smallest eigenvalues in the limit 
of large matrix size $N,$ one introduces the scaled variable $s=x/4N$
in terms of which 
the probability $E_{W,\beta}\left(0,s\right)$ 
that the interval $(0,s)$ is empty has been derived in Refs. \cite{TW2,Forrester} for
$\beta=2$ and  \cite{Forrester2} for $\beta=1.$ 
When perturbed, the above probabilities modify to 

\begin{equation}
E(0,s)=
 \frac{(b/a)^{\lambda/2}}
{2K_{\lambda}(\sqrt{ab})}
\int_0^{\infty} d\xi\xi^{\lambda-1}\exp
\left[-\frac{1}{2}(\frac{a}{\xi}+b\xi)\right]
E_{W,\beta}\left(s\sqrt{\frac{\xi} 
{\bar{\xi}}}\right) ,  \label{53}
\end{equation}
which shows how the smallest eigenvalue distribution is deformed in
the presence of a hyperbolic disorder. This probability becomes simple
when $M=N$ with the Marchenko-Pastur density diverging at the origin.
In this case the probability $E_{W,\beta}\left(0,s\right)$ assume simple 
exponential form particularly for $\beta=1.$

\section{Conclusion}

We have investigated the effect of superimposing an extra source 
of randomness governed by a generalized inverse Gaussian to the 
Gaussian fluctuations of the Wigner and the Wishart ensembles. 
The result is an ensemble of  random matrices ruled by the generalized 
hyperbolic distribution which contains as particular cases disordered
ensembles previously studied. The spectral density and 
short range statistics like spacing distribution show a transition
from Wigner-Dyson to Poisson  statistics, approaching 
universal critical statistics, namely semi-Poisson statistics, 
at an intermediate point of the transition. 
However, differently from the short range, long range 
statistics of the hyperbolic ensemble show a superpoissonian
behavior with  large fluctuations. The combination of semi-Poisson
at short range and super-Poisson at long range have
been observed in the non-ergodic embedded Gaussian models of 
many-body systems\cite{Bennet}. Therefore the present hyperbolic model
is more suited to give a simple way to model features associated
with the non-ergodicity of physical motivated ensembles.

This work is supported by a CAPES-COFECUB project and 
by the Brazilian agencies CNPq and FAPESP.

{\bf Figure Captions}

Fig. 1 The GIG distribution, Eq. (\ref{28}), is plotted for the 
indicated values of the parameters: in a) the parameters $a$
and $b$ are kept constants while in b) $\lambda=5$ and the ratio
$a/b$ is varied keeping $ab=4.$ 

Fig. 2 The transition of the density from the semi-circle to
a Gaussian-like for the indicated values of the parameters.

Fig. 3 The transition of the spacing $F(s)$ from Wigner-Dyson to
Poisson statistics is shown for the same values of the parameters of
Fig. 2.

Fig. 4 The nearest-neighbor distribution (NND) calculated with
the approximated equation (\ref {35}) for the same values of 
the parameters of Figs. 2 and 3.

Fig. 5 The number variance with small level of disorder is shown 
and compared with the GOE, the Poisson and the semi-Poisson
predictions.

Fig. 6 The density of the disordered Wishart matrices of size
M=80 and N=40 is compared with the Marchenko-Pastur density for
the indicated values of the parameters.

\end{document}